\documentclass[showpacs,preprintnumbers,amsmath,amssymb,twocolumn]{revtex4}
\usepackage{graphicx}
\usepackage{dcolumn}
\usepackage{bm}
\usepackage{amssymb}
\usepackage{array}
\usepackage{hhline}
\usepackage{longtable}

\begin{document}

\title{Determining global mean-first-passage time of random walks on Vicsek fractals\\ using  eigenvalues of Laplacian matrices}

\author{Zhongzhi Zhang$^{1,2}$}
\email{zhangzz@fudan.edu.cn}

\author{Bin Wu$^{1,2}$}

\author{Hongjuan Zhang$^{2,3}$}

\author{Shuigeng Zhou$^{1,2}$}
\email{sgzhou@fudan.edu.cn}

\author{Jihong Guan$^{4}$}
\email{jhguan@tongji.edu.cn}

\author{Zhigang Wang$^{5}$}

\affiliation {$^{1}$ School of Computer Science, Fudan University,
Shanghai 200433, China}

\affiliation {$^{2}$ Shanghai Key Lab of Intelligent Information
Processing, Fudan University, Shanghai 200433, China}

\affiliation {$^{3}$ Department of Mathematics, College of Science,
Shanghai University, Shanghai 200444, China}

\affiliation{$^{4}$ Department of Computer Science and Technology,
Tongji University, 4800 Cao'an Road, Shanghai 201804, China}

\affiliation{$^{5}$ Mahle Technologies Holding (China) Co., Ltd.,
Shanghai 201400, China}

\begin{abstract}

The family of Vicsek fractals is one of the most important and
frequently studied regular fractal classes, and it is of
considerable interest to understand the dynamical processes on this
treelike fractal family. In this paper, we investigate discrete
random walks on the Vicsek fractals, with the aim to obtain the
exact solutions to the global mean-first-passage time (GMFPT),
defined as the average of first-passage time (FPT) between two nodes
over the whole family of fractals. Based on the known connections
between FPTs, effective resistance, and the eigenvalues of graph
Laplacian, we determine implicitly the GMFPT of the Vicsek fractals,
which is corroborated by numerical results. The obtained closed-form
solution shows that the GMFPT approximately grows as a power-law
function with system size (number of all nodes), with the exponent
lies between 1 and 2. We then provide both the upper bound and lower
bound for GMFPT of general trees, and show that the leading behavior
of the upper bound is the square of system size and the dominating
scaling of the lower bound varies linearly with system size. We also
show that the upper bound can be achieved in linear chains and the
lower bound can be reached in star graphs. This study provides a
comprehensive understanding of random walks on the Vicsek fractals
and general treelike networks.

\end{abstract}

\pacs{05.40.Fb, 61.43.Hv, 89.75.Hc, 05.60.Cd}

\date{\today}
\maketitle

\section{Introduction}

Fractals are an important concept characterizing the features of
real systems, because they can model a broad range of objects in
nature and society~\cite{Ma82}. Over the past few decades, fractals
have attracted considerable interest from the physics
community~\cite{HaBe87,BeHa00}. Among numerous fractal classes, the
so-called regular fractals are an integral family of fractals.
Examples include the Sierpinski gasket~\cite{Si1915}, the Koch
snowflake~\cite{Ko1906}, the Vicsek fractals~\cite{Vi83}, and so on.
These structures have received much
attention~\cite{Ma82,HaBe87,BeHa00}, and continue to be an active
object of research~\cite{Fa03}. One of the main reasons for studying
regular fractals is that one can obtain explicit closed-form
solutions on a finite structure. Another justification is that
various problems intractable on Euclidean lattices become solvable
on regular fractals~\cite{ScScGi97}. On the other hand, the exact
solutions on regular fractals can provide useful insight different
from that given by the approximate solutions for random fractals.

A central issue, still debated, is to understand how the underlying
geometrical and structural features influence various dynamics
defined on complex systems, which has been considered to be an
important problem in many interdisciplinary fields, e.g. network
science~\cite{Ne03,BoLaMoChHw06,DoGoMe08}. Amongst a plethora of
fundamental dynamical processes, random walks are crucial to a lot
of branches of sciences and engineering and have appealed much
interest~\cite{HaBe87,MeKl00,MeKl04,BuCa05}. A basic quantity
relevant to random walks is first-passage time (FPT)~\cite{Re01},
which is the expected time to hit a target node for the first time
for a walker staring from a source node. It is a quantitative
indicator to characterize the transport efficiency, and carries much
information of random walks since many other quantities can be
expressed in terms of it. Thus, a growing number of studies have
been concentrated on this interesting
quantity~\cite{Mo69,NoRi04,CoBeMo05,SoRebe05,CoBeTeVoKl07,BaCaPa08,ZhZhZhYiGu09,TeBeVo09}.

In view of the significance of regular fractals and random-walk
dynamics, many authors have devoted their endeavors to study random
walk on regular fractals~\cite{HaRo08}, such as the Sierpinski
gasket~\cite{KaBa02PRE,KaBa02IJBC}, the
$T-$fractal~\cite{KaRe86,KaRe89,Ag08}, the Vicsek
fractals~\cite{Vi84,Vo09}, as well as the hierarchical lattice
fractals~\cite{BeOs79,ZhXiZhLiGu09}. The results of these
investigations unveiled many unusual and exotic phenomena of random
walks on regular fractals. But in the aspect of FPT, these studies
only addressed the mean of FPTs between part of the node pairs,
e.g., between a given node and all other
nodes~\cite{KaBa02PRE,KaBa02IJBC,Ag08,ZhXiZhLiGu09}, while the
scaling for the FPT averaged over all pairs of nodes, often called
global mean first-passage time (GMFPT), in the regular fractals is
still not well understood~\cite{CoBeTeVoKl07}, in spite that GMFPT
provides comprehensive information of random walks on fractals and
other media.

In this paper, we study analytically the discrete random walks on a
class of treelike fractals---Vicsek fractals, which are typical
candidates for exact mathematical fractals and have received
extensive
interest~\cite{WeGr85,BlJuKoFe03,WaLi92,StFeBl05,ZhZhChYiGu08}. We
determine exactly the GMFPT between two nodes over the whole fractal
family, which is verified by numerical results. The closed-form
formula for the GMFPT is achieved iteratively by using the advantage
of the specific construction of the Vicsek fractals. The obtained
explicit expression indicates that for large systems the GMFPT
increases algebraically with the size of the systems. In the second
part of this work, we provide the rigorous upper and lower bounds
for GMFPT as a function of system size for general treelike media.
We show that of all trees linear chains have the largest value of
GMFPT and the star graphs have the smallest GMFPT.

\section{Brief introduction to the Vicsek fractals}

\begin{figure}
\begin{center}
\includegraphics[width=.85\linewidth,trim=100 0 100 0]{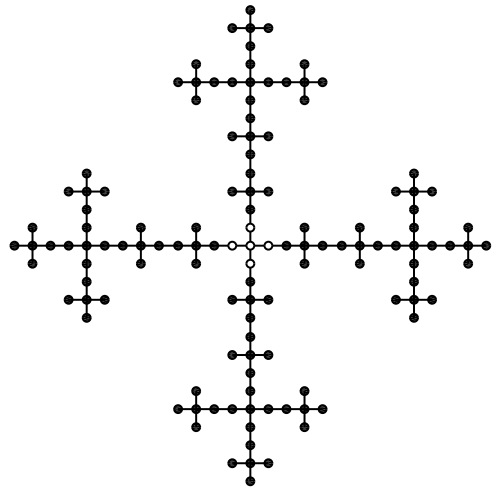}
\caption{Illustration of the first several iterative processes of a
particular Vicsek fractal $V_{4,3}$. The open circles denote the
starting structure $V_{4,1}$.} \label{net}
\end{center}
\end{figure}

The so-called Vicsek fractals are constructed in an iterative
way~\cite{Vi83,BlJuKoFe03}. Let $V_{f,g}$ ($f\geq 2$, $g \geq 1$)
denote the Vicsek fractals after $g$ iterations (generations). The
construction starts from ($g=1$) a star-like cluster consisting of
$f+1$ nodes arranged in a cross-wise pattern, where $f$ peripheral
nodes are connected to a central node. This corresponds to
$V_{f,1}$. For $g\geq 2$, $V_{f,g}$ is obtained from $V_{f, g-1}$.
To obtain $V_{f,2}$, we generate $f$ replicas of $V_{f,1}$ and
arrange them around the periphery of the original $V_{f,1}$, then we
connect the central structure by $f$ additional links to the corner
copy structure. These replication and connection steps are repeated
infinitely, with the needed Vicsek fractals obtained in the limit $g
\rightarrow \infty$, whose fractal dimension is $\ln (f+1)/\ln3$. In
Fig.~\ref{net}, we show schematically the structure of $V_{4,3}$.
According to the construction algorithm, at each step the number of
nodes in the systems increases by a factor of $f+1$, thus, we can
easily know that the total number of nodes (i.e., network order or
system size) of $V_{f,g}$ is $N_{g}= (f+1)^{g}$. Since the whole
family of Vicsek fractals has a treelike structure, the total number
of links in $V_{f,g}$ is $E_{g}= N_{g}-1=(f+1)^{g}-1$.

\section{GMFPT in the Vicsek fractals}

After introducing the Vicsek fractals $V_{f,g}$, we will continue to
study numerically and analytically random walks performed on them,
which is the primary topic of this present paper. The random-walk
model we study is a simple one. Assuming the time to be discrete, at
each time step, the walker (or particle) jumps uniformly from its
current location to one of its neighbors. The highly desirable
quantity related to random walks is the GMFPT starting from a source
point to a given target point, averaged over all node pairs of
source and target points.

The GMFPT can be obtained numerically but exactly via the
pseudoinverse~\cite{BeGr03,RaMi71} of the Laplacian matrix,
$\textbf{L}_g$, of $V_{f,g}$. The entries $L_{ij}^{g}$ of
$\textbf{L}_g$ are defined as follows: the off-diagonal element
$L_{ij}^{g}=-1$ if the pair of nodes $i$ and $j$ are linked to each
other, otherwise $L_{ij}^{g}=0$; while the diagonal entry
$L_{ii}^{(g)}=d_i$ (degree of node $i$). The pseudoinverse (denoted
by $\textbf{L}_g^\dagger$) of $\textbf{L}_g$ is a variant of its
inverse matrix and is defined to be
\begin{equation}\label{Pinverse01}
 \textbf{L}_g^\dagger=\left(\textbf{L}_g-\frac{\textbf{e}_g\textbf{e}_g^\top}{N_g}\right)^{-1}+\frac{\textbf{e}_g\textbf{e}_g^\top}{N_g}\,,
\end{equation}
where $\textbf{e}_g$ is the $N_g$-dimensional ``one" vector, i.e.,
$\textbf{e}_g=(1,1,\cdots,1)^\top$.

The FPT between any pair of nodes in $V_{f,g}$ can be expressed in
terms of the elements, $L_{ij}^{\dagger,g}$, of
$\textbf{L}_g^\dagger$. Let $T_{ij}(g)$ stand for the FPT for random
walks in $V_{f,g}$, starting from node $i$ to node $j$.
Then~\cite{CaAb08}
\begin{equation}\label{Hitting01}
 T_{ij}(g)=\sum_{n=1}^{N_g}\left(L_{in}^{\dagger,g}-L_{ij}^{\dagger,g}-L_{jn}^{\dagger,g}+L_{jj}^{\dagger,g}\right)L_{nn}^{g}\,,
\end{equation}
where $L_{nn}^{g}$ is the $n$th diagonal entry of $\textbf{L}_g$.
Thus, the sum, $T_{\rm sum}(g)$, for FPTs between all node pairs in
$V_{f,g}$ reads as
\begin{equation}\label{Hitting02}
T_{\rm sum}(g)=\sum_{i\neq j}\sum_{j=1}^{N_g}T_{ij}(g)\,,
\end{equation}
and the GMFPT, $\langle T \rangle_g$, is
\begin{equation}\label{Hitting03}
\langle T \rangle_g=\frac{T_{\rm
sum}(g)}{N_g(N_g-1)}=\frac{1}{N_g(N_g-1)}\sum_{i\neq
 j}\sum_{j=1}^{N_g}T_{ij}(g)\,.
\end{equation}

\begin{figure}
\begin{center}
\includegraphics[width=0.85\linewidth,trim=50 55 90 35]{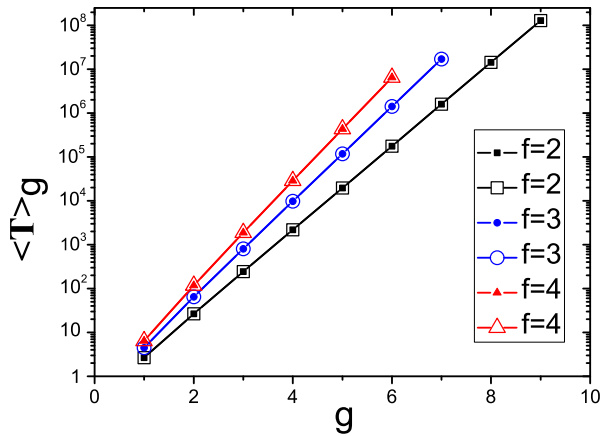}
\end{center}
\caption[kurzform]{\label{Time01} (Color online) Global
mean-first-passage time $\langle T \rangle_g$ as a function of the
iteration $g$ on a semilogarithmic scale for different parameter
$f$. The filled symbols are the numerical results obtained by direct
calculation from Eqs.~(\ref{Hitting01}) and (\ref{Hitting03}), while
the empty symbols correspond to the exact values from
Eq.~(\ref{Hitting10}), both of which are consistent with each
other.}
\end{figure}

Using Eqs.~(\ref{Hitting01}) and (\ref{Hitting03}), we can compute
directly the GMFPT $\langle T \rangle_g$ of the Vicsek fractals (see
Fig.~\ref{Time01}). From Fig.~\ref{Time01}, we can see that $\langle
T \rangle_g$ approximately grows exponentially in $g$. In other
words, $\langle T \rangle_g$ is a power-law function of network
order $N_g$ obeying the scaling as $\langle T \rangle_g \sim
(N_g)^{\theta}$ since $N_{g}= (f+1)^{g}$. It should be mentioned
that although the expression of Eq.~(\ref{Hitting03}) seems compact,
it requires computing the inversion of a matrix of order $N_{g}
\times N_{g}$ [see Eq.~(\ref{Pinverse01})], which make heavy demands
on time and computational resources for large networks. Thus, one
can calculate directly from Eq.~(\ref{Hitting03}) the GMFPT only for
the first iterations. On the other hand, by using the method of
pseudoinverse matrix it is difficult and even impossible to obtain
the leading behavior of the exponent $\theta$ characterizing the
random walks. It is thus of significant practical importance to seek
for a computationally cheaper method for computing the GMFPT.
Fortunately, the particular construction of the Vicsek fractals and
the connection~\cite{ChRaRuSm89,Te91} between effective resistance
and the FPTs for random walks allow us to calculate analytically the
GMFPT and the exponent $\theta$ to obtain rigorous solutions.

Below we will show how to avoid the computational complexity of
inverting a matrix. To this end, we view $V_{f,g}$ as resistor
networks~\cite{DoSn84} by considering each edge to be a unit
resistor. Let $R_{ij}(g)$ be the effective resistance between two
nodes $i$ and $j$ in the electrical networks obtained from
$V_{f,g}$. Then, according to the relation between FPTs and
effective resistance~\cite{ChRaRuSm89,Te91}, we have
\begin{equation}\label{Hitting04}
T_{ij}(g)+T_{ji}(g)=2\,E_g\,R_{ij}(g)\,.
\end{equation}
Therefore, Eq.~(\ref{Hitting02}) can be rewritten as
\begin{equation}\label{Hitting06}
T_{\rm sum}(g)=E_{g}\,\sum_{i\neq
 j}\sum_{j=1}^{N_g}R_{ij}(g)\,.
\end{equation}
Using the previously obtained results~\cite{GuMo96,ZhKlLu96}, the
sum term on the right-hand side of Eq.~(\ref{Hitting06}) denoted by
$R_{\rm sum}(g)$ can be recast as
\begin{equation}\label{Hitting08}
R_{\rm sum}(g)=\sum_{i\neq
 j}\sum_{j=1}^{N_g}R_{ij}(g)=2\,N_g\,\sum_{i=2}^{N_g}\frac{1}{\lambda_i^{(g)}}\,,
\end{equation}
where $\lambda_i^{(g)}$ ($i=2,\ldots, N_g$) are all the nonzero
eigenvalues of Laplacian matrix, $\textbf{L}_g$, of the Vicsek
fractals $V_{f,g}$. Then, we have
\begin{equation}\label{Hitting09}
 \langle T \rangle_g=2\,\sum_{i=2}^{N_g}\frac{1}{\lambda_i^{(g)}}\,.
\end{equation}
Having $\langle T \rangle_g$ in terms of the sum of the reciprocal
of all nonzero Laplacian eigenvalues, the next step is to determine
this sum.

The determination of all eigenvalues of $\textbf{L}_g$ can be
resolved by using the real-space decimation
method~\cite{DoAlBeKa83,Ra84}. Assuming that one has the eigenvalues
$\lambda_i^{(g)}$ ($\lambda_i^{(g)}\neq 0$) at generation $g$, then
the eigenvalues $\lambda_i^{(g+1)}$ of the next generation $g+1$ can
be obtained through the relation~\cite{JaWu92,JaWu94,BlFeJuKo04}
\begin{equation}\label{EigVal01}
\lambda_i^{(g+1)}(\lambda_i^{(g+1)}-3)(\lambda_i^{(g+1)}-f-1)=\lambda_i^{(g)}\,.
\end{equation}
By solving Eq.~(\ref{EigVal01}), each eigenvalue $\lambda_i^{(g)}$
($\lambda_i^{(g)}\neq 0$) at generation $g$ gives rise to three new
and different ones at generation $g+1$, denoted by
$\lambda_{i,1}^{(g+1)}$, $\lambda_{i,2}^{(g+1)}$, and
$\lambda_{i,3}^{(g+1)}$, respectively. Moreover, the newly generated
eigenvalues keep the degeneracy of their ancestors. Considering that
all the nonzero eigenvalues of $V_{f,1}$ are $\lambda_i^{(1)}=1$
($i=2,3,\ldots,f$) and $\lambda_{f+1}^{(1)}=f+1$, one can obtain all
nonzero eigenvalues $\lambda_i^{(g)}$ of $\textbf{L}_g$ by
iteratively solving Eq.~(\ref{EigVal01}) $g-1$ times.

It should be stressed that although we can provide $\lambda_i^{(g)}$
in a recursive way, it is difficult to write $\lambda_i^{(g)}$ in an
explicit formula. However, in what follows we will show that the
recursive solution to $\lambda_i^{(g)}$ allows to obtain a
closed-form expression for the sum of the reciprocal of all nonzero
eigenvalues of $\textbf{L}_g$, denoted by $\Lambda_g$. By definition
\begin{equation}\label{EigVal02}
\Lambda_g = \sum_{i=2}^{N_g} \frac{1}{\lambda_i^{(g)}}\,.
\end{equation}
A main goal of the following text is to explicitly determining this
sum.

Let $\Omega_g$ express the set of all the $N_g$ eigenvalues of
$\textbf{L}_g$, i.e., $\Omega_g=\{\lambda_1^{(g)},
\lambda_2^{(g)},\cdots,\lambda_{N_g}^{(g)}\}$, where the
distinctness of the elements has been ignored. Notice that all these
eigenvalues are either nondegenerate or
degenerate~\cite{BlFeJuKo04}. The set of the former is denoted by
$\Omega_g^{(1)}$, while the set of the latter is denoted by
$\Omega_g^{(2)}$. That is to say, $\Omega_g=\Omega_g^{(1)} \cup
\Omega_g^{(2)}$. $\Omega_g^{(1)}$ includes $0$, $f+1$ and other
eigenvalues generated by the ``seed" $\lambda_{f+1}^{(1)}=f+1$; and
$\Omega_g^{(2)}$ includes $1$ and other eigenvalues derived from
$1$. Furthermore, the degeneracy of the degenerate eigenvalues rests
with the generation at which they appeared at the first time. At a
given generation $j$, the degeneracy of eigenvalues $1$ is
$\Delta_j=(f-2)(f+1)^{j-1}+1$, a degeneracy that their descendants
keep. In what follows, for convenience we use $\Omega_g^{(1)}$ to
represent the nondegenerate eigenvalues of $\textbf{L}_g$ other than
$0$.


We now return to derive $\Lambda_g$, which can be evidently recast
as
\begin{equation}\label{EigVal03}
\Lambda_g = \sum_{\lambda_i^{(g)} \in
\Omega_g^{(1)}}\frac{1}{\lambda_i^{(g)}}+\sum_{\lambda_i^{(g)} \in
\Omega_g^{(2)}}\frac{1}{\lambda_i^{(g)}}\,.
\end{equation}
We denote the two sums on the right-hand side of
Eq.~(\ref{EigVal03}) by $\Lambda_g^{(1)}$, and $\Lambda_g^{(2)}$,
respectively. Below we will calculate the two quantities
$\Lambda_g^{(1)}$ and $\Lambda_g^{(2)}$.

We first calculate $\Lambda_g^{(1)}$. At the initial generation $1$,
there is only one nondegenerate eigenvalue $f+1$, which produces
three different nondegenerate eigenvalues at generation $2$. We call
these three eigenvalues the first-generation descendants of $f+1$,
which give rise to $3^2$ second-generation descendants of $f+1$ at
the third generation. Thus, at $i$th generation, $3^{i-1}$
$(i-1)$th-generation descendants of $f+1$ are produced. Since all
eigenvalues (degenerate or nondegenerate) which appeared at one
generation will still appear in all subsequent
generations~\cite{JaWu92,JaWu94,BlFeJuKo04}, we have
$\Omega_{g-1}^{(1)} \subset \Omega_g^{(1)}$. Hence, as noted above,
$\Omega_g^{(1)}$ consists of $f+1$ and all its offspring produced
after generation 1.

Let $\Gamma_{i}^{(1)}$ be the sum of the reciprocal of all the
$(i-1)$th-generation descendants of $f+1$. Then, $\Lambda_g^{(1)}$
can be rewritten in terms of $\Gamma_{i}^{(1)}$ as
\begin{equation}\label{EigVal04}
\Lambda_g^{(1)} = \sum_{i=0}^{g-1}\Gamma_{i}^{(1)}\,,
\end{equation}
where $\Gamma_{0}^{(1)}=\Lambda_1^{(1)}=1/(f+1)$.

Note that for each nonzero eigenvalue (degenerate or nondegenerate)
$\lambda_i^{(g)} \in \Omega_g$, Eq.~(\ref{EigVal01}) can be
rewritten in an alternative way as
\begin{equation}\label{EigVal05}
(\lambda_i^{(g+1)})^{3}-(f+4)(\lambda_i^{(g+1)})^{2}+3(f+1)\lambda_i^{(g+1)}-\lambda_i^{(g)}=0\,.
\end{equation}
According to the Vieta's formulas, the three roots (i.e.,
$\lambda_{i,1}^{(g+1)}$, $\lambda_{i,2}^{(g+1)}$, and
$\lambda_{i,3}^{(g+1)}$) of Eq.~(\ref{EigVal05}) satisfy the
following two relations: $\lambda_{i,1}^{(g+1)} \cdot
\lambda_{i,2}^{(g+1)}\cdot \lambda_{i,3}^{(g+1)}=\lambda_i^{(g)}$
and $\lambda_{i,1}^{(g+1)} \cdot
\lambda_{i,2}^{(g+1)}+\lambda_{i,1}^{(g+1)} \cdot
\lambda_{i,3}^{(g+1)}+\lambda_{i,2}^{(g+1)} \cdot
\lambda_{i,3}^{(g+1)}=3(f+1)$. Thus,
$1/\lambda_{i,1}^{(g+1)}+1/\lambda_{i,2}^{(g+1)}+1/\lambda_{i,3}^{(g+1)}=3(f+1)/\lambda_i^{(g)}$.
Based on the results obtained above, we have
\begin{equation}\label{EigVal06}
\Gamma_g^{(1)} =\sum_{\lambda_i^{(g)} \in \Omega_g^{(1)}\backslash
\Omega_{g-1}^{(1)}}\frac{3(f+1)}{\lambda_i^{(g)}}=3(f+1)\Gamma_{g-1}^{(1)}\,,
\end{equation}
which together with the initial condition $\Gamma_{0}^{(1)}=1/(f+1)$
leads to $\Gamma_{g}^{(1)}=3^g(f+1)^{g-1}$. Inserting this result
into Eq.~(\ref{EigVal04}), we get
\begin{equation}\label{EigVal07}
\Lambda_g^{(1)} =
\sum_{i=0}^{g-1}\left[3^i(f+1)^{i-1}\right]=\frac{1}{f+1}\frac{3^{g}(f+1)^{g}-1}{3f+2}\,.
\end{equation}

After obtaining $\Lambda_g^{(1)}$, all that is left to find an
expression for $\Lambda_g$ is to evaluate $\Lambda_g^{(2)}$. For
each eigenvalue 1, applying an approach similar to that used above,
we can compute the sum of the reciprocal of its $(i-1)$th-generation
descendants, which we represent by $\Upsilon_{i}^{(2)}$. After some
simple algebra, we obtain $\Upsilon_{i}^{(2)}=3^{i}(f+1)^{i}$ ($ 0
\leq i \leq g-1$), where $\Upsilon_{0}^{(2)}=1$ express the
reciprocal of the ``seed'' eigenvalue $1$ itself. It has been shown
that~\cite{JaWu92,JaWu94,BlFeJuKo04} in the Vicsek fractals
$V_{f,g}$, the degeneracy of eigenvalues $1$ is
$\Delta_g=(f-2)(f+1)^{g-1}+1$, and the degeneracy of each of its
$i$th-generation ($0 \leq i\leq g-1$) offspring is
$\Delta_{g-i}=(f-2)(f+1)^{g-1-i}+1$. Then, the quantity
$\Lambda_g^{(2)}$ is evaluated as follows:
\begin{eqnarray}\label{EigVal08}
\Lambda_g^{(2)} &=& \sum_{i=0}^{g-1}\left(\Delta_{g-i}\cdot
\Upsilon_{i}^{(2)}\right)\nonumber \\
&=&\frac{(f-2)(f+1)^{g-1}(3^{g}-1)}{2}+\frac{3^{g}(f+1)^{g}-1}{3f+2}\,,
\end{eqnarray}

Plugging Eqs.~(\ref{EigVal07}) and~(\ref{EigVal08}) into
Eq.~(\ref{EigVal03}) yields
\begin{equation}\label{EigVal09}
\Lambda_g =
\frac{(f-2)(f+1)^{g-1}(3^{g}-1)}{2}+\frac{f+2}{f+1}\frac{3^{g}(f+1)^{g}-1}{3f+2}\,.
\end{equation}
Using the  relation $\langle T \rangle_g=2\Lambda_g$, we have
\begin{equation}\label{Hitting10}
\langle T
\rangle_g=(f-2)(f+1)^{g-1}(3^{g}-1)+\frac{2(f+2)}{f+1}\frac{3^{g}(f+1)^{g}-1}{3f+2}\,.
\end{equation}
We have confirmed this closed-form expression for $\langle T
\rangle_g$ against direct computation from Eqs.~(\ref{Hitting01})
and~(\ref{Hitting03}). For all range of $g$ and different $f$, they
completely agree with each other, which shows that the analytical
formula provided by Eq.~(\ref{Hitting10}) is right.
Figure~\ref{Time01} shows the comparison between the numerical and
predicted results, with the latter plotted by the full expression
for the sum in Eq.~(\ref{Hitting10}).

We can also support the validity of Eq.~(\ref{Hitting10}) by using
another method. In fact, the correctness of Eq.~(\ref{Hitting10})
depends on all the nonzero Laplacian eigenvalues, the exactness for
derivation of which can be established according to the relation
between the Laplacian eigenvalues and the number of spanning trees
of a graph. It has been established that the number of spanning
tress on a connected graph $G$ with order $N$, $N_{\rm{st}}(G)$, is
related to all its nonzero Laplacian eigenvalues $\lambda_i$
(assuming $\lambda_1=0$ and $\lambda_i\neq 0$ for $i=2,\cdots, N$),
obeying the relation
$N_{\rm{st}}(G)=\frac{1}{N}\prod_{i=2}^{N}\lambda_i$~\cite{TzWu00}.
Since the Vicsek fractals $V_{f,g}$ are trees, the product of all
nonzero Laplacian eigenvalues for $V_{f,g}$, denoted by $\Theta_g$,
should equal $N_g$, which can be corroborated by the following
argument. By definition,
$\Theta_g=\Theta_g^{(1)}\cdot\Theta_g^{(2)}$, where $\Theta_g^{(i)}$
($i=1,2$) is the product of Laplacian eigenvalues in
$\Omega_g^{(i)}$. Applying the Vieta's formulae, we can easily
obtained the product of the $i$th-order ($0 \leq i \leq g-1$)
offspring of the ``seed'' eigenvalue $f+1$ is $f+1$, which is
independent of $i$. Then $\Theta_g^{(1)}=(f+1)^g$. Similarly, we
have $\Theta_g^{(2)}=1$. Hence, $\Theta_g=(f+1)^g=N_g$, which proves
the correctness of the computation on the Laplacian eigenvalues for
$V_{f,g}$.

We proceed to show how to represent GMFPT, $\langle T\rangle_g$, as
a function of the network order $N_g$, with the aim to obtain the
relation between these two quantities. Recalling $N_{g}= (f+1)^{g}$,
we have $3^g=(N_g)^{\ln 3/\ln (f+1)}$ that enables one to write
$\langle T \rangle_g$ in the following form:
\begin{eqnarray}\label{Hitting11}
\langle T \rangle_g=&\quad&\frac{f-2}{f+1}N_{g}[(N_g)^{\ln 3/\ln
(f+1)}-1]\nonumber\\
&+&\frac{2(f+2)}{(f+1)(3f+2)}[(N_g)^{1+\ln 3/\ln (f+1)}-1]\,.
\end{eqnarray}

\begin{figure}
\begin{center}
\includegraphics[width=0.85\linewidth,trim=50 40 80 35]{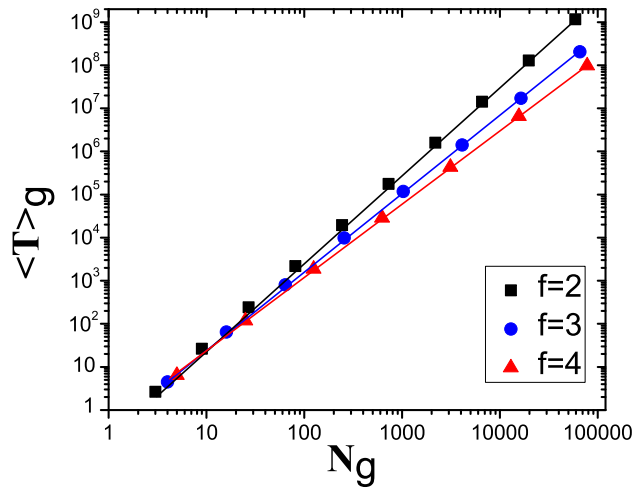}
\end{center}
\caption[kurzform]{\label{Time02} (Color online) Global mean
first-passage time $\langle T \rangle_g$ versus the network order
$N_g$ on a log-log scale. The filled symbols described the analytic
results shown in Eq.~(\ref{Hitting10}). The solid lines represent
the corresponding leading scaling given by Eq.~(\ref{Hitting12}).}
\end{figure}

Equation~(\ref{Hitting11}) unveils the explicit dependence relation
of GMFPT on network order $N_g$ and parameter $f$. For large
systems, i.e., $N_g\rightarrow \infty$, we have following expression
for the dominating term of $\langle T \rangle_g$:
\begin{eqnarray}\label{Hitting12}
\langle T \rangle_g &\sim& \frac{f(3f-2)}{(f+1)(3f+2)}(N_g)^{1+\ln
3/\ln
(f+1)}\nonumber \\
&=&\frac{f(3f-2)}{(f+1)(3f+2)}(N_g)^{\theta}\nonumber \\
&=&\frac{f(3f-2)}{(f+1)(3f+2)}(N_g)^{2/\tilde{d}}\,
\end{eqnarray}
where $\tilde{d}=2\ln(f+1)/\ln(3f+3)$ is the spectral dimension of
the Viskek fractals~\cite{BlFeJuKo04}. Thus, in the large limit of
$g$, the GMFPT grows approximately as a power-law function of
network order $N_g$ with the exponent $\theta=1+\ln 3/\ln (f+1)$
being a decreasing function of $f$. It is easy to see that the
exponent $\theta$ is larger than 1 but not greater than 2.
Particularly, when $f=2$, $\theta$ reduces to 2, which is the
highest one reported thus far. In fact, 2 is largest exponent for
GMFPT of random walks defined on treelike media, the rigorous proof
of which will be given in the next section. In addition, it should
be mentioned that the obtained superlinear dependence of GMFPT on
the network order is in contrast with the other scalings previously
observed for other media, e.g., linear scaling for the Apollonian
networks~\cite{HuXuWuWa06} and the pseudofractal scale-free
web~\cite{Bobe05}, a logarithmic correction to the linear dependence
for small-world trees~\cite{BaCaPa08,TeBeVo09}. Figure~\ref{Time02}
shows how the GMFPT scales with the network
order for various parameter $f$. 

\section{Bounds for GMFPT in trees}

In Sect. III, we have shown that the GMFPT in the Vicsek fractals
scales as a power-law function of network order. Previous studies
exhibited that GMFPT in other trees may depend on network order $N$
following different scalings. For example, in the deterministic
uniform recursive trees, their GMFPT varies with network order $N$
as $N\ln N$~\cite{ZhQiZhGaGu10}; in the $T-$fractal, the GMFPT grows
as $N^{1+\ln 2/\ln 3}$~\cite{ZhLiZhWuGu09}. These show that in
different trees, the GMFPT obeys different dependence relation on
network order. Then, some natural questions arise: what are the
upper and lower bounds for GMFPT in general trees? In which trees
are these bounds reached?

As a matter of fact, the above questions are equivalent to find the
upper and lower bounds for the total effective resistance, $R_{\rm
sum}$, as defined similarly by Eq.~(\ref{Hitting08}). One can prove
with ease using various
methods~\cite{EnJaSn76,Pl84,Lolo03,ZhZhWaSh07,GhBoSa08} that for
trees with order $N$, the upper and lower bounds for $R_{\rm sum}$
are
\begin{equation}\label{UppBoun}
R_{\rm sum}^{\rm Upp} = \frac{N(N-1)(N+1)}{3}\,
\end{equation}
and
\begin{equation}\label{LowBoun}
R_{\rm sum}^{\rm Low} = 2(N-1)^{2}\,,
\end{equation}
respectively.

The upper bound can be only reached for the tree that is exactly a
linear chain (a path), which has two nodes with degree 1 at both
ends of the chain and $N-2$ nodes with degree 2 in the
middle~\cite{ZhZhWaSh07}. Actually, this linear chain is one of the
particular Viscek fractals corresponding to $f=2$. The result
provided by Eq.~(\ref{UppBoun}) is compatible with that of the
Vicsek fractals corresponding to $f=2$. As for the lower bound, it
can be only achieved when the tree is a star
graph~\cite{Pl84,Lolo03,GhBoSa08}, consisting of one central node
and $N-1$ leaf nodes. All these leaf nodes are linked to the central
node, and there is no edge between the leaf nodes.

From Eqs.~(\ref{UppBoun}) and~(\ref{LowBoun}), we can easily obtain
that the upper and lower bounds for GMFPT are
\begin{equation}\label{UppGMFPT}
\langle T \rangle^{\rm Upp} = \frac{(N-1)(N+1)}{3}\,
\end{equation}
and
\begin{equation}\label{LowGMFPT}
\langle T \rangle^{\rm Low} =\frac{ 2(N-1)^{2}}{N}\,,
\end{equation}
respectively. Thus, the GMFPT $\langle T \rangle$ for general trees
satisfies the relation $\langle T \rangle^{\rm Upp} \leq \langle T
\rangle \leq \langle T \rangle^{\rm Low}$. For large trees (i.e., $N
\rightarrow \infty$), the leading scalings for $\langle T
\rangle^{\rm Upp}$ and $\langle T \rangle^{\rm Low}$ change
separately with network order $N$ as $\langle T \rangle^{\rm Upp}
\sim N$ and $\langle T \rangle^{\rm Low} \sim N^2$, implying that
the scaling for the GMFPT in any tree must lie between linear
scaling and square of network order.  It is very obvious that the
upper bound is much larger than the lower bound, the reasons for
which lie with the underlying structures of the corresponding
graphs: the linear chain is homogeneous, while the star graph is
heterogeneous.

In the star graphs, the central node has a very large degree, and
thus plays a crucial role in keeping the whole graph together. When
the random-walk process is performed in the star graphs, the walker
has a tendency to migrate toward the central node, through which it
jumps to the target nodes. Therefore, the efficiency of random walks
is very high in the star graphs, the linear scaling of the GMFPT
with $N$ is the best we can see~\cite{GhBoSa08}. Notice that the
same scaling has been previously observed for complete
graphs~\cite{Bobe05}. In fact, the star graphs can be obtained from
the complete graph with the same order by whittling down complete
graphs, i.e., by the judiciously removing edges from complete graphs
leaving only one node with $N-1$ connections, in order that the
walker in the star graphs can find the destination nodes as easily
as in the complete graphs.

On the contrary, in the linear chains all nodes are homogenous. When
the walker starting from the source point to find the target node
far away from the staring point, it must traverse all nodes between
the starting point and the destination node. This makes the traverse
time much longer than in the star graphs.

Finally, we should stress that although the star graphs are extreme
of heterogenous media, they are very instructive to understand the
dynamics of random walks on other heterogeneous graphs, especially
scale-free networks~\cite{BaAl99}, which are ubiquitous in real
natural and social systems~\cite{AlBa02,DoMe02}. Previous studies
have shown that random walks in scale-free networks are very
efficient~\cite{Bobe05,ZhQiZhXiGu09,ZhGuXiQiZh09,ZhZhXiChLiGu09,AgBu09,ZhLiGoZhGuLi09,ZhXiZhLiGu09},
the roots of which is actually can be heuristically explained as
above. The large-degree nodes in scale-free networks play a similar
role as that of the central node in the star graphs, making the
GMFPT very small.

\section{Conclusions}

We have studied the discrete random walks on the family of Vicsek
fractals, which includes the linear chains as a particular case.
Using the connection between the FPTs and the Laplacian eigenvalues
for general graphs, we have computed the GMFPT averaged over all
pairs of nodes in the fractals and obtained explicit solutions to
the GMFPT. The obtained closed-form formula shows that in the limit
of infinite network order $N$, the GMFPT $ \langle T \rangle $ grows
approximately as a power-law function of $N$: $\langle T \rangle
\sim N^{1+\ln 3/\ln (f+1)}$. We have also provided rigorous bounds
on the network order dependence of the GMFPT in general treelike
networks. We showed that the upper and lower bounds can be achieved
in linear chains and star graphs, respectively. Our study sheds
useful insights into the random-walk process occurring on treelike
media.

\subsection*{Acknowledgments}

We would like to thank Yuan Lin for assistance. This work was
supported by the National Natural Science Foundation of China under
Grants No. 60704044, No. 60873040, and No. 60873070; the National
Basic Research Program of China under Grant No. 2007CB310806; the
Shanghai Leading Academic Discipline Project No. B114, and the
Program for New Century Excellent Talents in University of China
(Grant No. NCET-06-0376). B.W. also acknowledges the support by
Fudan's Undergraduate Research Opportunities Program, and H.J.Z.
acknowledges the support by Shanghai Key Laboratory of Intelligent
Information Processing, China. Grant No. IIPL-09-017.

\end{document}